\documentclass[manuscript]{acmart}
\AtBeginDocument{%
  }

\setcopyright{acmlicensed}
\copyrightyear{2018}
\acmYear{2018}
\acmDOI{XXXXXXX.XXXXXXX}
\acmConference[Conference acronym 'XX]{Make sure to enter the correct
  conference title from your rights confirmation email}{June 03--05,
  2018}{Woodstock, NY}
\acmISBN{978-1-4503-XXXX-X/2018/06}

\usepackage{multirow}
\usepackage{csquotes}
\usepackage{xcolor}
\usepackage{soul}
\sethlcolor{lightgray}
\usepackage{makecell}
\usepackage{subcaption}

\begin{document}

\title{Synthetic Heuristic Evaluation: A Comparison between AI- and Human-Powered Usability Evaluation}

\author{Ruican Zhong}
\orcid{0009-0004-7169-0675}
\email{rzhong98@uw.edu}
\affiliation{%
  \institution{University of Washington}
  \city{Seattle}
  \state{WA}
  \country{USA}}
\author{David W. McDonald}
\orcid{0000-0001-5882-828X}
\email{dwmcphd@gmail.com}
\affiliation{%
  \institution{University of Washington}
  \city{Seattle}
  \state{WA}
  \country{USA}}

\author{Gary Hsieh}
\orcid{0000-0002-9460-2568}
\email{garyhs@uw.edu}
\affiliation{%
  \institution{University of Washington}
  \city{Seattle}
  \state{WA}
  \country{USA}}



\begin{abstract}
Usability evaluation is crucial in human-centered design but can be costly, requiring expert time and user compensation. In this work, we developed a method for synthetic heuristic evaluation using multimodal LLMs' ability to analyze images and provide design feedback. Comparing our synthetic evaluations to those by experienced UX practitioners across two apps, we found our evaluation identified 73\% and 77\% of usability issues, which exceeded the performance of 5 experienced human evaluators (57\% and 63\%). Compared to human evaluators, the synthetic evaluation's performance maintained consistent performance across tasks and excelled in detecting layout issues, highlighting potential attentional and perceptual strengths of synthetic evaluation. However, synthetic evaluation struggled with recognizing some UI components and design conventions, as well as identifying across screen violations. Additionally, testing synthetic evaluations over time and accounts revealed stable performance. Overall, our work highlights the performance differences between human and LLM-driven evaluations, informing the design of synthetic heuristic evaluations.
\end{abstract}



\keywords{Do, Not, Us, This, Code, Put, the, Correct, Terms, for,
  Your, Paper}

\received{20 February 2007}
\received[revised]{12 March 2009}
\received[accepted]{5 June 2009}
\maketitle
\section{Introduction}
Usability evaluation, as a way to understand how users perceive the usability of an interface and to identify usability problems, is a key part of the human-centered design process~\cite{dingli2014intelligent}. However, conducting usability testing can be costly. Performing a 5-participant study can cost up \$10k to \$50k~\cite{usabilitycost}, and may take up 11 to 27 hours of time~\cite{userevalstats}. To reduce the overall cost, prior work has explored the use of simulations to automate the evaluation of user interfaces. Yet reviews~\cite{ivory2001state,namoun2021review} of existing approaches revealed that there has not been a system that fully automates the evaluation process using user interaction or perceptions of the interface. Further, existing approaches' results are often limited in consistency and accuracy~\cite{balbo1995automatic,bowers1996weblint,al1999kaldi,hammontree1992integrated,fosco2020predicting,lee2020guicomp, uehling1995user}. %Overall, there remains a gap in considering how to reduce the cost of running user evaluations . 

The development of LLMs (large language models) shows a possibility of overcoming these limitations and advancing usability evaluation. Indeed, prior literature suggests that LLMs are able to provide feedback and critiques on user interfaces~\cite{park2022social, park2023generative, duan2024generating}. Duan et al. developed an AI-driven Figma plugin that provides design feedback given JSON descriptions of a UI~\cite{duan2024generating}. Wu et al. also explored using multimodal LLMs to provide tips about UI designs~\cite{wu2024uiclip}. However, while this recent research has demonstrated the feasibility of using LLMs to produce design suggestions, prior work did not specifically study their use for usability evaluation, where validity and reliability are important to ensure the rigor of the process~\cite{hughes1999rigor}. %This is still an unexplored area of AI-powered synthetic heuristic evaluation.
Despite multimodal LLMs' capabilities, their stochastic nature~\cite{ouyang2023llm} and hallucination issues have been noted~\cite{mckenna2023sources, radhakrishnan2023question, morris2024prompting}, where they have been shown to produce inaccurate or inconsistent outputs. Without a systematic evaluation, it is unclear if multimodal LLMs can be used to effectively perform usability evaluation. In addition, by comparing synthetic evaluation outputs against human evaluations, we may better explore potential differences between the two and advance our understanding of how LLMs ``think'' about UI design and feedback. 

%are there notable differences between LLM-powered evaluations and human evaluations? 

%Understanding the features of synthetic evaluation outputs and comparing them against human outputs will enable us to better understand how LLMs ``think'' about UI design and feedback. 

Thus, in our study, we conducted a systematic study of the use of multimodal LLMs for heuristic evaluations \cite{nielsen1990heuristic} -- one of the most widely used usability methods. We study how the performance of our synthetic heuristic evaluation compares to that of expert human evaluators, and how reliable our synthetic heuristic evaluation is across repeated prompting. In addition, we contrasted the performance of three off-the-shelf LLMs (GPT-4, Gemini-1.5-pro, Claude 3.5 Sonnet) in conducting synthetic heuristic evaluation. We ask the following research questions:
% And more recently, Vision Language Models (VLM), also known as multimodal LLMs, have been introduced to analyze and interpret images directly~\cite{hudson2019gqa,saikh2022scienceqa,zeng2022socratic}, and may be useful in evaluating user interfaces.  
\begin{itemize}
    \item RQ1: Can LLMs be prompted to perform synthetic heuristic evaluations?
    \item RQ2: How do synthetic heuristic evaluations compare to human heuristic evaluations?
    \item RQ3: How reliable is synthetic heuristic evaluation across repeated prompting?
    \item RQ4: How does the performance of off-the-shelf LLMs compare to one another?
\end{itemize}

We first developed a process to instruct GPT-4 to conduct synthetic heuristic evaluations on interfaces, using Nielsen's 10 heuristics~\cite{nielsen1992finding}. We found that the synthetic evaluation identified 73\% and 77\% of the usability issues in two apps, which was more than the number of usability issues identified by the aggregation of 5 expert evaluators (57\% and 63\%). We also tested and showed the reliability of our approach across two accounts over a three-month period. In addition, the comparison between the LLM platforms revealed that GPT-4 had the best performance amongst the three tested in conducting synthetic heuristic evaluation.

Beyond performance and reliability, our work also highlights several differences between LLM and human evaluators. Our analyses indicate that synthetic evaluation's performance remained consistent while expert evaluators' performance decreased across evaluation tasks. And the synthetic evaluation outperformed the expert evaluators in identifying issues related to consistency in layout, indicating its ability to pick up on detailed aesthetic violations. However, synthetic evaluation sometimes had trouble recognizing and understanding the design of some UI elements. Additionally, our comparison indicated that synthetic evaluation is limited in utilizing information from multiple screens and identifying across screen violations. These results point to the strengths and limitations of LLM-powered synthetic evaluation, especially in comparison to human evaluations. These characteristics can inform the design of LLM-powered synthetic evaluation, which is another gap in prior research.

Overall, we contribute:
\begin{itemize}
    \item An effective way of prompting LLMs to perform synthetic heuristic evaluation
    \item An empirical evaluation demonstrating the efficacy of using LLM to perform heuristic evaluation
    \item Insights about how LLM-powered synthetic evaluation may differ from those done by humans
    \item A reliability test of the performance of LLM-powered heuristic evaluation over time
    \item A comparison of the performance of synthetic heuristic evaluation across multiple off-the-shelf LLMs
    % \item A systematic way to compare the performances of off-the-shelf LLMs
\end{itemize}

% Propose an effective way of prompting to perform synthetic HE; reliable and stable prompt
\section{Related Works}
\subsection{Heuristic Evaluation}
\label{relatedwork:HE}
Heuristic evaluation is a commonly used formative usability evaluation technique that compares user interfaces against a given set of design heuristics -- a set of principles commonly used to ensure the usability of the interface, pointing out any violations~\cite{nielsen1992finding}. To conduct a heuristic evaluation, an evaluator is provided with a list of heuristics and is tasked to review the interface design and note down any issues violating the provided heuristics 
~\cite{nielsen1990heuristic}. One of the most common sets of heuristics is Nielsen's 10 heuristics. It was created by Jakob Nielsen and Molich in the 1990s~\cite{molich1990improving, nielsen1992finding}. It includes the following ten heuristics: \textit{Visibility of system status, Match between system and the real world, User control and freedom, Consistency and standards, Error prevention, Recognition rather than recall, Flexibility and efficiency of use, Aesthetic and minimalist design, Help users recognize, diagnose and recover from errors, Help and documentation.}

For each violation found, evaluators are also asked to rate the severity of the usability issues from 0 to 4 to help prioritize the most important issues to fix. 0 refers to \textit{``I don't agree that this is a usability problem at all.''} 1 refers to \textit{``Cosmetic problem only: need not be fixed unless extra time is available on project.''} 2 refers to \textit{``Minor usability problem: fixing this should be given low priority.''} 3 refers to \textit{``Major usability problem: important to fix, so should be given high priority.''} 4 refers to \textit{``Usability catastrophe: imperative to fix this before product can be released.''} 

One of the original studies of heuristic evaluation found that individual heuristic evaluators were able to find between 20\% to 50\% of the usability problems in the interfaces they evaluate, and that aggregating evaluations from 5 evaluators can result in 55\%-90\% coverage ~\cite{nielsen1990heuristic}. A follow-up study showed that evaluators' expertise matter, and that non-experts are less effective at identifying usability issues than usability specialists ~\cite{nielsen1992finding}. 3-5 usability specials were able to find 74\%-87\% of the violations, while 5 non-experts were able to uncover 50\% of the violations. This resulted in the often repeated recommendation that heuristic evaluation should be conducted by three to five independent evaluations to provide sufficient coverage.  

To evaluate the performance of heuristic evaluation, prior works first developed a master set, encompassing all potential heuristic issues with the interface. Then the researchers would compare each evaluator's outputs against that set and identify the percentage of issues covered by that evaluator. To understand multiple evaluators' overall performances, the researchers would first coalesce all the evaluators' identified issues and calculate the percentage of issues covered in the master set. This approach of comparing against a ground truth set of problems ensures that each evaluator's performance is evaluated objectively and equally\mbox{~\cite{nielsen1990heuristic,mankoff2003heuristic,langevin2021heuristic}}.

In our work, we chose heuristic evaluation as the user evaluation technique to automate using LLM due to several reasons. First, heuristic evaluation is a flexible technique that has been shown to be effective for evaluating web and mobile interfaces ~\cite{akula2021critical,azizi2021usability, fung2016heuristic, branco2022usability, al2016heuristic}. By adapting the set of 10 original heuristics by Nielsen et al., heuristic evaluation can also be extended to a variety of novel interfaces and interactions ~\cite{mankoff2003heuristic,langevin2021heuristic,magoulas2003integrating}. Second, heuristic evaluation does not require the evaluators to be users or have experience with the interface evaluated ~\cite{nielsen1992finding}. This prevents the need to have domain-specialized LLMs. Third, heuristic evaluation can be done by inspecting screenshots and does not require interactions with the actual interface~\cite{nnheuristic}. Analyses of screenshots is now possible with multimodal LLMs~\cite{hudson2019gqa,saikh2022scienceqa,zeng2022socratic}, while more complex interactions are not yet directly possible and would require additional engineering.  

\subsection{Automated Usability Testing}
To reduce the cost of conducting usability testing with actual users~\cite{usabilitycost,userevalstats}, prior works explored how to automate the testing process~\cite{ivory2001state}. Yet, according to Ivory \& Hearst's survey~\cite{ivory2001state}, there are two major limitations to these systems. First, none of the existing tools achieved full automation. For instance, some prior research only automated the portion of the process that occurs after user data collection is complete. They would interpret recordings of user interactions with interfaces to identify usability problems~\cite{rauterberg1995novice,cugini1999visvip,webtrends}. However, studies to collect user data are still required. And since user evaluation sessions are costly to conduct~\cite{usabilitycost,userevalstats}, this type of system does not address the fundamental cost of collecting user data. Some other existing systems attempted to simulate usage traces to represent user interactions~\cite{chi2000scent, helfrich1999quip, olsen1988interface}. However, these prior approaches only simulated user behavior through a rule-based approach. The mechanisms behind were rather mechanical and do not capture user interactions realistically, indicating a lack of full automation support as well.    

Another issue with existing approaches is that they were unable to capture the qualitative insights and subjective information in usability testing sessions. While some existing systems record user data~\cite{rauterberg1995novice,cugini1999visvip,webtrends}, they only take notes of the quantitative measures (e.g., task completion time~\cite{al1999kaldi,hammontree1992integrated}, eye-tracking~\cite{fosco2020predicting,lee2020guicomp}, guideline violations~\cite{uehling1995user}, etc.). However, these quantitative measures are difficult to interpret~\cite{baumeister2000comparison,glenn1992development}. Thus, these existing approaches ``do not capture important qualitative and subjective information (such as user preferences and misconceptions) that can only be unveiled via usability testing, Heuristic evaluation and other standard inquiry method~\cite{ivory2001state}.'' Some other existing approaches would use the codebase of the application to understand how the app is implemented and conduct evaluations accordingly. For instance, they would consider whether a specific label exists for a text field. Yet, they also fail to consider users’ perspectives on how to interact with the interfaces~\cite{balbo1995automatic,bowers1996weblint}. 

Additionally, existing approaches' results are often limited in consistency and accuracy. A recent review of the current automated evaluation tools shows that all the aforementioned limitations persist (e.g., using quantitative measures and only producing a score output). And Namoun et al. found that there is a high variance in the performance of the tools, indicating that purely using quantitative measures may lead to inconsistent results~\cite{namoun2021review}.

\subsection{Use of Generative AI for Design Feedback}
The recent development of generative AI has introduced the possibility of advancing prior limitations in automated evaluation. LLMs have been trained on a huge amount of information (e.g., scientific papers, discussion forums, etc.), including feedback about interactions with user interfaces. These information might reflect aspects of how humans think about a specific problem and make decisions. Additionally, the recent development of multimodal LLMs has now enabled the direct processing and analysis of images. And prior works have shown how multimodal LLMs could generate summarization of image content~\cite{zeng2022socratic}. Such simulation of cognition and perception implies that multimodal LLMs might be able to simulate human interactions with interfaces and provide design feedback accordingly~\cite{schmidt2024simulating}. 

Indeed, some prior works have suggested the potential to do so. A commercialized tool, Synthetic Users\footnote{https://www.syntheticusers.com/}, boasts the use of generative AI to create user personas. It uses these personas to reflect on product ideas and simulate how different audiences would react to an idea, indicating how to optimize user journey, prioritize product roadmap, and increase product conversion rate. While it helps generate feedback on product ideas, it does not focus on critiquing the design of user interfaces. Zhang et al. developed LlamaTouch, an LLM-powered mobile UI task automation approach that reads a screenshot of a mobile app and identifies the possible next step/interaction with the screenshot to achieve a given user task~\cite{zhang2024llamatouch}. Similarly, the Claude 3.5 Sonnet model released a ``computer use'' feature, enabling the model to walk through the process of conducting a user task~\cite{claude-sonnet}. Yet these tools only identify the right steps to achieve a user task. They do not point out any design violations heuristic along the way and do not provide feedback on design. Overall, existing systems fail to capture qualitative insights about the UIs, which is a crucial part of usability testing.

Duan et al. recently developed an LLM-powered Figma plugin that generates design feedback based on a set of guidelines (the guidelines include heuristics from Nielsen's)~\cite{duan2024generating}. They also developed a UI critique dataset that enabled LLMs to provide design feedback regarding a bounded region on a screenshot~\cite{duan2024uicrit}. Yet there are some limitations to their work. First, design critiques and heuristic evaluation are fundamentally different. Heuristic evaluation is a usability testing methodology where each usability issue identified should be a violation of the given set of heuristics. The description should clearly articulate why and how the design violates the heuristics, requiring more reasoning ability. In contrast, design critiques are more broadly and vaguely defined, where any reflection about the design could be a design critique. Thus, LLMs' ability to perform design critique may not carry over to heuristic evaluation. Additionally, while Duan et al.~\cite{duan2024generating} conducted user studies to show the possibility of using LLMs to provide design critiques using JSON descriptions of images, they did not systematically demonstrate LLMs' efficacy of doing so (as discussed in \mbox{\autoref{relatedwork:HE}}). Instead, they focused on understanding UX designers' perceptions of these outputs.

Wu et al. also created UIClip~\cite{wu2024uiclip}, a fine-tuned machine learning model that can generate UI design suggestions. Again, their work only demonstrated LLM's ability to generate design feedback but did not capture the aspect of heuristic evaluation. Additionally, their work only evaluated LLMs' performance quantitatively, using metrics such as quality and relevancy. They did not analyze the qualitative features of LLM outputs.
% Overall, they did not investigate the characteristics of LLM outputs and did not systematically demonstrate in which aspects LLMs' outputs differ from the human evaluation.
% In addition, the inputs to generative AI in Duan et al.'s tool are text-based descriptions of the interfaces instead of actual images of the interfaces.
% As discussed, multimodal LLMs' ability to analyze screenshots of interfaces \mbox{~\cite{hudson2019gqa,saikh2022scienceqa,zeng2022socratic}} can help address this. We could now do away with the added step of having to first convert interfaces to text-based representation. 
In our study, we address these gaps in prior works. We investigate the use of multimodal LLMs to analyze images and provide human-like usability evaluation, focusing on Nielsen's 10 heuristics. We provide a systematic data analysis that shows how LLM-powered heuristic evaluation and human evaluation are similar and different. This informs how we should think about synthetic evaluations when using them.

\subsection{Testing the Reliability of Generative AI}
Recently, scholars have come to acknowledge the need to consider the reliability of LLM models, which has a strong impact on the replication of results~\cite{morris2024prompting}. Due to how LLMs are trained and how they generate output, the LLM-powered heuristic evaluation results may differ every single time, even though the inputs (screenshots of the apps) are the same, especially as time progresses and as the models are updated. Yet, none of the existing works exploring the use of generative AI to provide design feedback accounted for the stochastic nature of LLMs~\cite{duan2024generating,duan2024uicrit,wu2024uiclip}. While some prior research benchmarked LLMs' performances across multiple platforms~\cite{duan2024generating}, they have not considered how time could play a factor in understanding the efficacy of LLMs. In our work, we tested LLMs' performances both across multiple platforms and over a three month period of time.
% prompting considered harmful [TODO: CITE]
% to demonstrate the reliability of synthetic evaluation.

% In addition, LLMs [reliability]
% Moreoever, none of these existing research sought to address and test the realiability of LLMs. 

% Thus, in this study, we investigate whether multimodal LLMs can analyze images and provide human-like usability evaluation, focusing on Nielsen’s 10 heuristics. 

% consistency with repeated prompting seems unnecessary, as consistency can be controlled through temperature and top-k settings.

\section{Synthetic Evaluation Design}
\label{section:prompt}

\begin{table*}[]
\centering
\small
\caption{This table shows our iterative prompting process. It includes the prompts we tried, the example responses when using a prompt, and the issues we noticed when using that prompt. The highlighted portion in each prompt is the change compared to the previous row's version. }
\begin{tabular}{|p{7cm}|p{4cm}|p{4cm}|p{0cm}|}
% \vfill
% \toprule
\hline
\textbf{Prompt}                            & \textbf{Example Response}     &
\textbf{Issue with Example Response}
\\ 
\hline
\texttt{[User scenario: The user is trying to complete the initial set up process on a rental application, and encounters the following screens.] Perform heuristic evaluation using Nielsen's 10 heuristics based on the screenshot given. For each heuristic, identify at least 2 problems. }  & 

\textit{Recognition rather than recall}: \textit{The icons at the bottom are intuitive, aiding in recognition over recall for navigating the app.} &

The identified issues are not actual violations of the heuristics. Instead, it is only providing a confirmation that there was no issue.

\\

% \hline
% \texttt{Given the screenshot provided, perform a heuristic evaluation using Neilsen's 10 heuristics. \hl{For each heuristic, provide a rationale for why this is an issue.}} &

% &

% \\ 

\hline
\texttt{[User scenario: The user is trying to complete the initial set up process on a rental application, and encounters the following screens.] Given the screenshot provided, perform a heuristic evaluation using Nielsen's 10 heuristics. For each heuristic, identify at least 2 problems. \hl{Identify all heuristic issues, provide a rationale for why this is an issue, a severity rating (0-4), and a reason for the severity rating. Be as specific as possible about where the heuristic fails.}} &

When evaluating a rental application:
\textit{Match between system and the real world: The term ``amenities'' might be too vague or technical for some users.}

\textit{}

&
The LLM can clearly articulate the issues with the current given screenshot, but does not consider heuristic violations across multiple screens at all.

\\
\hline
\texttt{[User scenario: The user is trying to complete the initial set up process on a rental application, and encounters the following screens.] Given the screenshots provided, perform a heuristic evaluation using Nielsen's 10 heuristics. \hl{(The screenshots are given in the order that they show up in the application, so consider the interaction across the screens.)} For each heuristic, identify at least 2 problems. Identify all heuristic issues, provide a rationale for why this is an issue, give a severity rating (0-4) and reason for the severity rating. Be as specific as possible about where the heuristics fail.}  & 
\textit{Visibility of system status: There is no indication of the progress within the setup process. Users might not know how far they are from completing the initial setup.} \ldots \ldots \textit{Aesthetic and minimalist design: The ``Tour req \hl{(cut off because of token limit)}}

&
Each heuristic violation identified is clear. The LLM can provide detailed descriptions about some across screen usability issues. But the outputs were often cut off in the middle because token limit was reached. 
\\

\hline
\texttt{[User scenario: The user is trying to complete the initial set up process on a rental application, and encounters the following screens.] Given the screenshots provided, perform a heuristic evaluation using the \hl{\{Choose one of the following: first 5 / second 5}\} of Nielsen's 10 heuristics. (The screenshots are given in the order that they show up in the application, so consider the interaction across the screens.) For each heuristic, identify at least 2 problems. Identify all heuristic issues, provide a rationale for why this is an issue, give a severity rating (0-4) and reason for the severity rating. Be as specific as possible about where the heuristics fail.}

& 
\textit{Visibility of System Status: There is no indication of the progress within the setup process. Users might not know how far they are from completing the initial setup.} \ldots\ldots\ldots
\textit{Aesthetic and Minimalist Design: The ``Tour requested'' message box is given visual prominence, while the actionable element, ``View Messages,'' is much smaller and less visually emphasized. Similarly, the pricing information is small compared to other less relevant elements.}

&
N/A. The LLM is able to identify heuristic violations consistently and can provide detailed descriptions about some across screen usability issues. The two outputs generated were complete.

\\
\hline
\end{tabular}

\label{table:prompts}
\end{table*}
We used GPT-4 in our study because it is one of the widely used state-of-the-art LLMs~\cite{openai, openaimodel}, and existing work has shown its ability to generate useful feedback on UIs based on a set of design guidelines~\cite{duan2024generating}. Since prompting is a key step to ensure GPT-4's output quality and consistency~\cite{wu2023scattershot}, we iterated on our prompts multiple times until GPT-4 generated reasonable results. Throughout this process, we noted three key challenges, including prompting the LLM to identify actual heuristic violations, to consider across screen violations, and to bypass the output token limit. 

The first notable problem is that the LLM had trouble identifying actual heuristic violations when we just used off-the-shelf LLM with a short prompt describing our task. As shown in the first row in \autoref{table:prompts}, we provided a description of the user task, and prompted LLM to simply do a heuristic evaluation. This type of prompting did not lead to reasonable results. Most of the issues described were not heuristic violations at all. Instead, the LLM said that the UI components were\textit{ ``easy to understand,'' ``intuitive,''} and \textit{``follow Nielsen's 10 heuristics.''} Thus, the LLM did not understand that the meaning of heuristic evaluation is to point out violations of the heuristics. To address this issue, we fine-tuned the prompt using the chain-of-thought technique~\cite{wei2022chain, zhang2022automatic, yu2023towards}. Specifically, we broke down the task of heuristic evaluation by asking the LLM to identify an issue, clarify the reason why there is an issue, and provide a severity rating accordingly (as shown in the second row of \autoref{table:prompts}). With this more targeted and detailed prompting, the LLM started identifying actual heuristic violations, such as \textit{``The tab `Collections' is ambiguous and does not clearly indicate what it contains or how it relates to the 'Review' process,''} and \textit{``The `X' to close the tip is very close to the device's status icons, which could lead to accidental closure.''} These are actually violations of Nielsen’s 10 heuristics upon investigation. At this stage, we noticed that the synthetic evaluation can already achieve a high coverage of usability issues. 

However, we did notice that the synthetic evaluation did not find any across screen heuristic evaluations. We speculate this is because multimodal LLM has a limited context window. To address this, we iterated our prompt with targeted instructions again to consider information from multiple screens. We added the phrase  \textit{``The screenshots are given in the order that they show up in the application, so consider the interaction across the screens.''} (as shown in the third row in \autoref{table:prompts}) We observed that while the LLM did not capture all across screen issues, it was able to report a reasonable amount, including issues such as \textit{``progress indication on the top of the screen is inconsistent''} and \textit{``the button placement may not be consistent as did on the previous screen.''} 
 
The last issue we encountered was the output token limit. We observed that the generated outputs were often cut off in the middle and did not cover the full 10 heuristics because they identified many violations and provided a rationale for all of them. To address this problem, we changed our prompting so that the LLM considers the first 5 heuristics in one exchange. Then, we provide the LLM with the same screenshot but ask it to consider the second 5 heuristics. The prompts we used are shown in the fourth row in \autoref{table:prompts}.

\section{Study 1: Synthetic Evaluation Performance Study} 
In the first experiment, we designed a study to understand whether LLMs can be prompted to perform synthetic heuristic evaluation and how its performance compares to human evaluations. 

\subsection{Procedure}
\begin{figure}[!tbp]
  \centering
  \begin{minipage}[b]{0.47\textwidth}
    \vspace{2pt}
        \centering
        \begin{subfigure}[t]{0.5\textwidth}
            \centering
            \includegraphics[width=\linewidth]{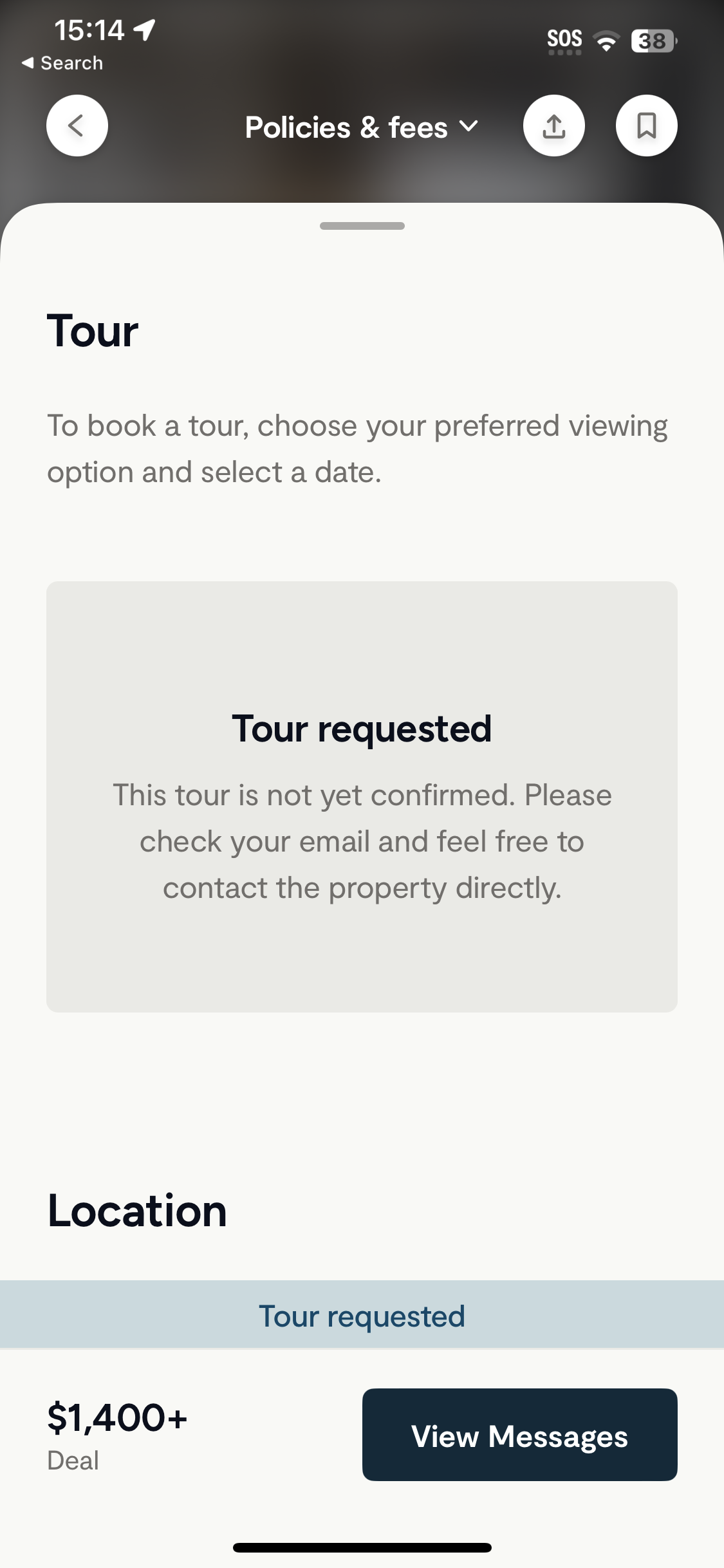}
            \caption{}
            \label{fig:example1}
        \end{subfigure}%
        ~ 
        \begin{subfigure}[t]{0.5\textwidth}
            \centering
            \includegraphics[width=\linewidth]{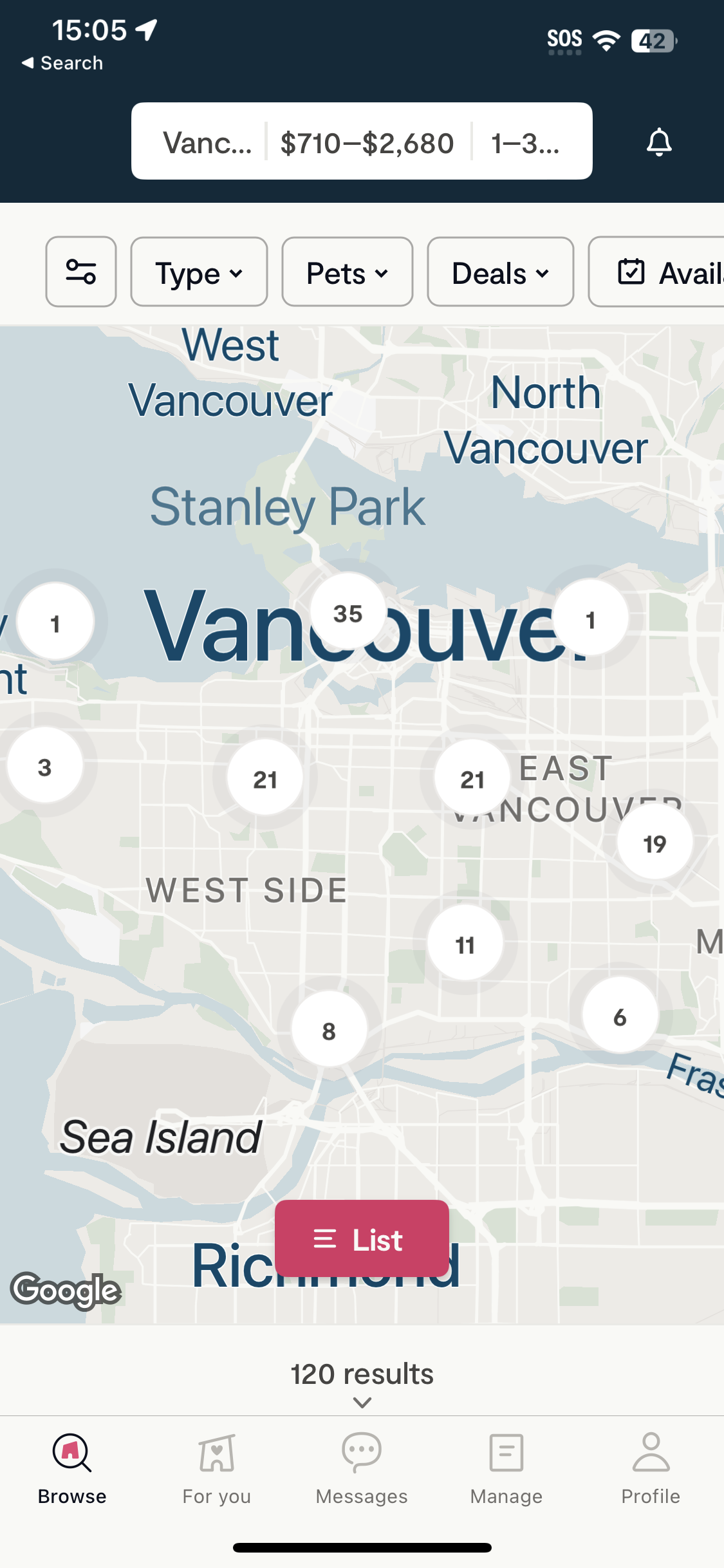}
            \caption{}
            \label{fig:example3}
        \end{subfigure}
        \caption{These two figures are example screens in the rental app.}

    % \label{fig:example1}
  \label{fig:rental}
  \end{minipage}
  \hfill
  \begin{minipage}[b]{0.47\textwidth}
    \vspace{2pt}
        \begin{subfigure}[t]{0.5\textwidth}
            \centering
            \includegraphics[width=\linewidth]{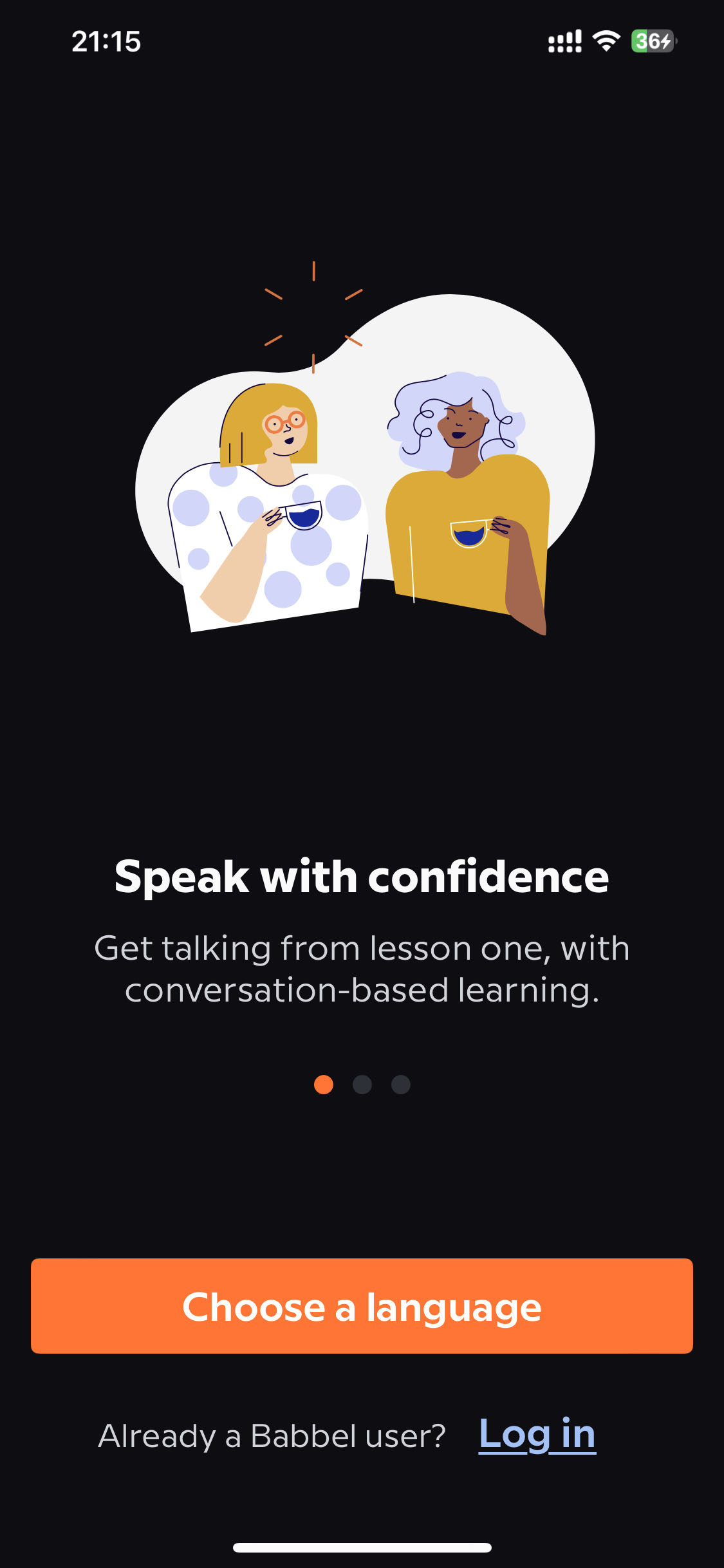}
            \caption{}
            \label{fig:example2}
        \end{subfigure}%
        ~ 
        \begin{subfigure}[t]{0.5\textwidth}
            \centering
            \includegraphics[width=\linewidth]{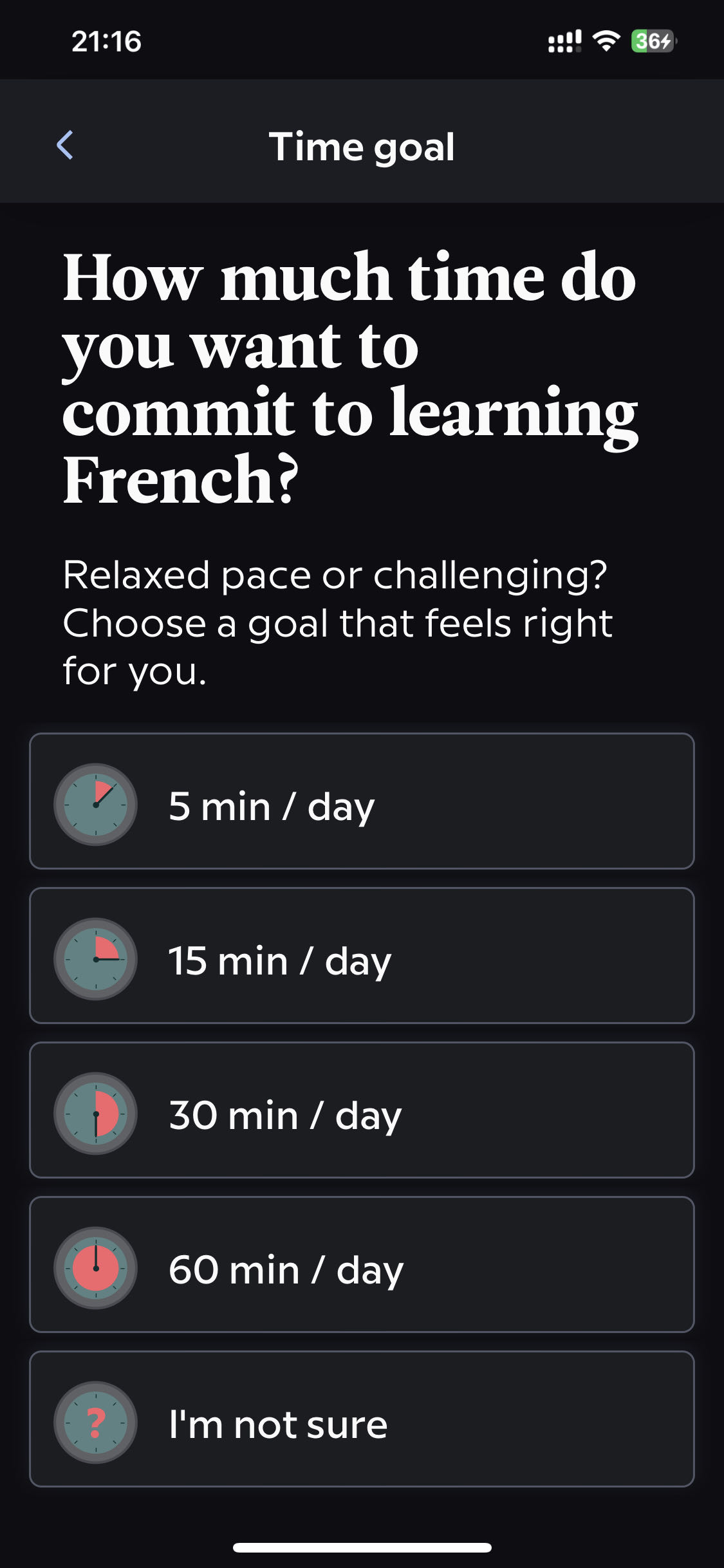}
            \caption{}
            \label{fig:example4}
        \end{subfigure}
        % \label{fig:language}
        \caption{These two figures are example screens in the language learning app.}
    \label{fig:language}
  \end{minipage}
\end{figure}

We selected two mobile applications to evaluate the performance and reliability of LLMs in synthetic heuristic evaluations: a rental application and a language learning application. These apps are representative of two popular categories of mobile apps: Lifestyle and Education~\cite{appcategory}. For both apps, we took 3-9 screenshots (i.e., \autoref{fig:rental}, \autoref{fig:language}) of a user performing a set of tasks. For the rental app, the tasks were \textit{``set up rental search preferences,''} \textit{``search for an apartment using a criteria,''} \textit{``explore the details about an apartment listing,''} and \textit{``book a visiting tour.''}
For the language learning app, the tasks were \textit{``set up learning goal for the French language,''} \textit{``explore the home screen to find a learning module,''} \textit{``experience a French learning lesson,''} and \textit{``set up a learning schedule.''} Our study protocol was approved by the IRB. 
% We presented both human evaluators and GPT-4 with the same set of screenshots. 

In our study, we generated three sets of evaluation results for both mobile applications using the aforementioned screenshots and user tasks: synthetic evaluation set, expert evaluator set, and the master set.

% \begin{figure*}[t!]
%     \centering
%     \begin{subfigure}[t]{0.24\textwidth}
%         \centering
%         \includegraphics[width=\linewidth]{image/example2.PNG}
%         \caption{}
%     \end{subfigure}%
%     ~ 
%     \begin{subfigure}[t]{0.24\textwidth}
%         \centering
%         \includegraphics[width=\linewidth]{image/example4.jpg}
%         \caption{}
%     \end{subfigure}
%     \caption{These two figures are example screens in the language learning app.}
% \end{figure*}
% \begin{figure}[!tbp]
%   \centering
%   \begin{minipage}[b]{0.4\textwidth}
%     \vspace{2pt}
%     \includegraphics[width=.7\linewidth]{image/example1.PNG}
%     \caption{This figure shows an example screen in the rental app.}
%     \label{fig:example1}
%   \end{minipage}
%   \hfill
%   \begin{minipage}[b]{0.4\textwidth}
%     \vspace{2pt}
%     \includegraphics[width=.7\linewidth]{image/example2.PNG}
%     \caption{This figure shows an example screen in the language learning app.}
%     \label{fig:example2}
%   \end{minipage}
% \end{figure}

% For the rental app, we asked both human evaluators and GPT-4 to experience the following tasks:  This application focuses on interactions that are relatively linear, where users simply follow along to set up searches and filter results. 

% In contrast, the evaluation tasks of the language learning app contain more dynamic interactions. 

\subsubsection{Synthetic evaluation set (GPT-4))}
\label{synthetic-set}
Using the process outlined in \autoref{section:prompt}, we prompted GPT-4 to complete a synthetic evaluation of both the rental app and the language learning app. Then, we grouped the results by the usability issues identified and analyzed them. We will refer to this set as the synthetic evaluation set.

\subsubsection{Expert evaluator set}
We recruited participants on UpWork\footnote{https://www.upwork.com/}, a job search platform for experienced freelancers, to complete heuristic evaluations of the rental app and the language learning app. We used ``UX Designer'' and ``heuristic evaluation'' as keywords to filter and invite workers from the platform. After workers accepted the invitation, we sent them a screener survey to select participants based on their prior work experience in UX design and heuristic evaluation. In total, we recruited 10 participants, 5 were randomly assigned to evaluate the rental app and 5 assigned to the language learning app. 70\% of the participants had completed more than 7 work projects in usability testing, and 80\% of the participants had completed more than 4 work projects related to heuristic evaluations. Their average self-reported familiarity with heuristic evaluation is 4.3 on a scale of 1-5 where 1 indicates ``Not familiar at all'' and 5 indicates ``Extremely familiar''.

We provided our expert evaluators with instructions similar to those we used to prompt GPT-4. Again, we first introduced the context of the mobile app and the usability tasks that the user was performing. Then, we asked the participants to conduct the heuristic evaluation based on Nielsen's 10 heuristics. Here is an example instruction used for one of the tasks in the rental app evaluation:
\begin{quote}
\begin{displayquote}
    \small
    \texttt{[The user is trying to complete the initial setup process on a rental application and encounters the following screens.] Follow the instructions below, and perform a heuristic evaluation using Nielsen's 10 heuristics.
    Please provide the severity rating and rationale for each heuristic issue identified, and provide recommendations to address each issue identified.}
\end{displayquote}
\end{quote}

Each participant was asked to complete this heuristic evaluation independently and asynchronously in an online templated document where we had the instructions and the screenshots of the apps. All participants sent a corresponding message upon completion of the work and received a \$30 compensation. After collecting all the data, we aggregated the results by the usability issues identified. 

\subsubsection{Master set}
\label{section:masterset}
The master set of heuristic issues represents the complete list of heuristic violations of the two apps uncovered in our study. This complete set would be used as a benchmark to demonstrate the performance of the LLM and the human evaluators. To curate the master set, we first combined all heuristic violations found by our expert evaluators and all issues found by GPT-4. In addition, we had five local research assistants (trained in conducting heuristic evaluation) conduct independent evaluations of the rental app and language learning app. The research assistants all had 4+ years of experience with HCI research, and on average, they had conducted 10+ heuristic evaluation sessions. They then came together to discuss the issues they identified, and those were added to the compiled master set. Integrating the additional set from the trained research assistants, our master set becomes more comprehensive.

In this combined set, we found duplicated issues, where the descriptions are phrased differently, but the descriptions refer to the same issue. To code this, one researcher reviewed all the issues in the set, then grouped and labeled all the issues that were duplicates. Two other researchers then came together to discuss these codes and reached a conclusion. We discovered that synthetic evaluation had found 8 duplicate issues for the rental app and 9 duplicates for the language learning app. Expert evaluators found 0 duplicate issues for both apps. After removing the duplicate issues in the set, we found that there were 182 unique issues identified for the rental app and 168 unique issues identified for the language learning app. 

Next, we coded the severity rating of each issue in the master set. We needed to do this because LLM's and each human evaluator's severity rating may be subjective and thus not as consistent, we decided to manually code the severity of each issue in the master set. The first author coded each heuristic violation with Nielsen's 10 heuristics and assigned a corresponding severity rating to it. 3 researchers then came together to reach a consensus on these ratings. Then, the first author re-coded the severity rating of the issues in the set accordingly. We found that there were 49 issues for the rental app (43 identified by synthetic evaluation, 6 identified by expert evaluators) and 55 issues for the language learning app (52 identified by synthetic evaluation, 3 identified by expert evaluators) that were of severity 0 (\textit{``I don't agree that this is a usability problem at all.''} as specified by Nielsen's). See \autoref{table:severity}. In total, there were 133 heuristic violations (severity $\neq$ 0) found for the rental app, and 113 violations (severity $\neq$ 0) found for the language learning app.

% Throughout this process, we found that there were 49 issues for the rental app (43 identified by synthetic evaluation, 6 identified by expert evaluators) and 58 issues for the language learning app (52 identified by synthetic evaluation, 3 identified by expert evaluators) that were of severity 0 (\textit{``I don't agree that this is a usability problem at all.''} as specified by Nielsen's). We excluded these issues in our finalized master set, which contained a total of 133 issues for the rental app and 112 for the language learning app. 

\subsection{Analyses}
To compare the performance of synthetic evaluation with expert evaluation, we first contrasted the number of issues identified by synthetic evaluation within the master set to those identified by the expert evaluators. We then analyzed the data by coded heuristic and severity rating in the master set to observe any significant performance differences. In addition to the quantitative analysis, we conducted a series of qualitative analyses on the reported issues across sets to gain further insights into the potential differences between synthetic evaluation and human evaluation.

\subsection{Results}
\label{section:result1}

\begin{figure}[!tbp]
  \centering
  \begin{minipage}[b]{0.4\textwidth}
    \includegraphics[width=\textwidth]{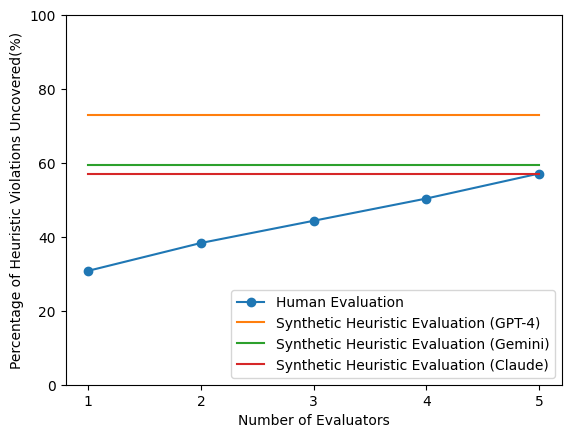}
    \caption{This figure shows expert evaluators' performance against synthetic heuristic evaluations' (GPT-4, Gemini, Claude) performance for the rental app}
    \label{fig:app1}
  \end{minipage}
  \hfill
  \begin{minipage}[b]{0.4\textwidth}
    \includegraphics[width=\textwidth]{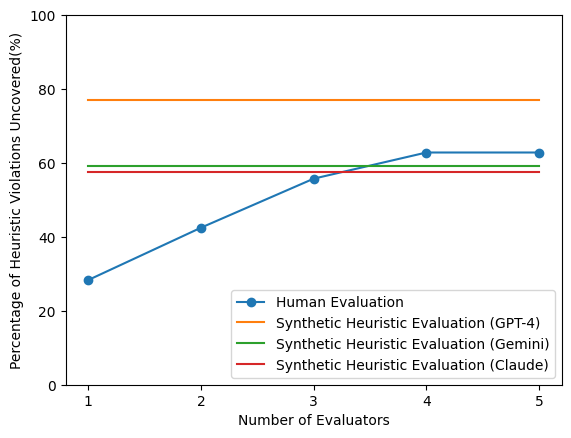}
    \caption{This figure shows expert evaluators' performance against synthetic evaluations' (GPT-4, Gemini, Claude) performance for the language learning app}
    \label{fig:app2}
  \end{minipage}
\end{figure}

% \usepackage{multirow}
% \usepackage{multirow}
% \usepackage{multirow}

% \usepackage{multirow}
\begin{table}[]
\caption [Short Heading]{\protect This table shows the following information by heuristics for both the rental app and the language learning app: the number of issues the average human evaluator detected, the coverage the average human evaluator achieved, the number of issues the synthetic evaluations (GPT-4, Gemini, Claude) detected, and the coverage the synthetic evaluations (GPT-4, Gemini, Claude) achieved. \autoref{section:result1} compares performance of GPT-4 against that of human evaluators and \autoref{reliability-result} compares the performance of GPT-4 against that of Gemini and Claude.}
\begin{tabular}{l|l|r|r|r|r}
Heuristic &
  App &
  \multicolumn{1}{l|}{\begin{tabular}[c]{@{}l@{}}Percentage of \\Issues by \\ Synthetic \\Evaluation\\ (GPT-4)\end{tabular}} &
  \multicolumn{1}{l|}{\begin{tabular}[c]{@{}l@{}}Percentage of \\Issues by \\ 5 Expert \\Evaluators\end{tabular}} &
  \multicolumn{1}{l|}{\begin{tabular}[c]{@{}l@{}}Percentage of \\Issues by \\ Synthetic \\Evaluation\\ (Gemini)\end{tabular}} &
  \multicolumn{1}{l}{\begin{tabular}[c]{@{}l@{}}Percentage of \\Issues by\\ Synthetic \\Evaluation\\ (Claude)\end{tabular}} \\ \hline
\multirow{2}{*}{Visibility of system status} &
  Rental &
  70\% (21/30) &
  57\% (17/30) &
  53\% (16/30) &
  53\% (16/30) \\ \cline{2-6} 
 &
  Language &
  87\% (26/30) &
  60\% (18/30) &
  70\% (21/30) &
  63\% (19/30) \\ \hline
\multirow{2}{*}{\begin{tabular}[c]{@{}l@{}}Match between system and \\ the real world\end{tabular}} &
  Rental &
  61\% (14/23) &
  48\% (11/23) &
  48\% (11/23) &
  43\% (10/23) \\ \cline{2-6} 
 &
  Language &
  74\% (14/19) &
  58\% (11/19) &
  47\%\space\space\space(9/19) &
  47\%\space\space\space(9/19) \\ \hline
\multirow{2}{*}{User control and freedom} &
  Rental &
  63\%\space\space\space\space\space(5/8) &
  88\%\space\space\space\space\space(7/8) &
  50\%\space\space\space\space\space(4/8) &
  38\%\space\space\space\space\space(3/8) \\ \cline{2-6} 
 &
  Language &
  55\%\space\space\space(6/11) &
  73\%\space\space\space(8/11) &
  45\%\space\space\space(5/11) &
  36\%\space\space\space(4/11) \\ \hline
\multirow{2}{*}{Consistency and standards} &
  Rental &
  46\%\space\space\space(6/13) &
  54\%\space\space\space(7/13) &
  31\%\space\space\space(4/13) &
  23\%\space\space\space(3/13) \\ \cline{2-6} 
 &
  Language &
  40\%\space\space\space(4/10) &
  60\%\space\space\space(6/10) &
  20\%\space\space\space(2/10) &
  20\%\space\space\space(2/10) \\ \hline
\multirow{2}{*}{Error prevention} &
  Rental &
  100\%\space\space\space\space\space(1/1) &
  0\%\space\space\space\space\space(0/1) &
  100\%\space\space\space\space\space(1/1) &
  100\%\space\space\space\space\space(1/1) \\ \cline{2-6} 
 &
  Language &
  100\%\space\space\space\space\space(2/2) &
  100\%\space\space\space\space\space(2/2) &
  100\%\space\space\space\space\space(2/2) &
  100\%\space\space\space\space\space(2/2) \\ \hline
\multirow{2}{*}{Recognition rather than recall} &
  Rental &
  100\%\space\space\space\space\space(1/1) &
  0\%\space\space\space\space\space(0/1) &
  100\%\space\space\space\space\space(1/1) &
  100\%\space\space\space\space\space(1/1) \\ \cline{2-6} 
 &
  Language &
  50\%\space\space\space\space\space(1/2) &
  100\%\space\space\space\space\space(2/2) &
  0\%\space\space\space\space\space(0/2) &
  0\%\space\space\space\space\space(0/2) \\ \hline
\multirow{2}{*}{Flexibility and efficiency of use} &
  Rental &
  82\%\space\space\space(9/11) &
  45\%\space\space\space(5/11) &
  36\%\space\space\space(4/11) &
  64\%\space\space\space(7/11) \\ \cline{2-6} 
 &
  Language &
  100\%\space\space\space\space\space(6/6) &
  50\%\space\space\space\space\space(3/6) &
  67\%\space\space\space\space\space(4/6) &
  83\%\space\space\space\space\space(5/6) \\ \hline
\multirow{2}{*}{Aesthetic and minimalistic design} &
  Rental &
  86\% (18/21) &
  38\%\space\space\space(8/21) &
  81\% (17/21) &
  67\% (14/21) \\ \cline{2-6} 
 &
  Language &
  74\% (14/19) &
  42\%\space\space\space(8/19) &
  58\% (11/19) &
  58\% (11/19) \\ \hline
\multirow{2}{*}{\begin{tabular}[c]{@{}l@{}}Help users recognize, diagnose\\ and recover from errors\end{tabular}} &
  Rental &
  67\%\space\space\space\space\space(2/3) &
  67\%\space\space\space\space\space(2/3) &
  33\%\space\space\space\space\space(1/3) &
  33\%\space\space\space\space\space(1/3) \\ \cline{2-6} 
 &
  Language &
  100\%\space\space\space\space\space(2/2) &
  100\%\space\space\space\space\space(2/2) &
  50\%\space\space\space\space\space(1/2) &
  50\%\space\space\space\space\space(1/2) \\ \hline
\multirow{2}{*}{Help and documentation} &
  Rental &
  91\% (20/22) &
  86\% (19/22) &
  91\% (20/22) &
  91\% (20/22) \\ \cline{2-6} 
 &
  Language &
  100\% (12/12) &
  92\% (11/12) &
  100\% (12/12) &
  100\% (12/12) \\ \hline
\end{tabular}
\label{table:heuristic}
\end{table}

\begin{table}[]

\caption [Short Heading]{\protect This table shows the following about the combination of the rental app and language learning app across severity rating for synthetic evaluation and the human evaluation: number of issues detected, the coverage achieved. Specifically, the comparison between GPT-4, Gemini, and Claude for the synthetic evaluation are demonstrated here and are explained in detail in \autoref{reliability-result}.}
\begin{tabular}{r|rl|r|r|r}
\multicolumn{1}{l|}{Severity} &
  \multicolumn{2}{l|}{\begin{tabular}[c]{@{}l@{}}Percentage of Issues by \\ Synthetic Evaluation\\ (GPT-4)\end{tabular}} &
  \multicolumn{1}{l|}{\begin{tabular}[c]{@{}l@{}}Percentage of Issues by \\ 5 Expert Evaluators\end{tabular}} &
  \multicolumn{1}{l|}{\begin{tabular}[c]{@{}l@{}}Percentage of Issues by\\ Synthetic Evaluation\\ (Gemini)\end{tabular}} &
  \multicolumn{1}{l}{\begin{tabular}[c]{@{}l@{}}Percentage of Issues by\\ Synthetic Evaluation\\ (Claude)\end{tabular}} \\ \hline
0          & \multicolumn{2}{r|}{65\%   (95/147)}   & 6\%   (9/147)  & 93\% (137/147)    & 84\% (123/147)    \\ \hline\hline
\textbf{1} & \multicolumn{2}{r|}{77\% (110/142)}    & 60\% (85/142)  & 65\%   (92/142)   & 63\%   (89/142)   \\ \hline
\textbf{2} & \multicolumn{2}{r|}{68\%     (40/59)}  & 59\%   (35/59) & 53\%     (31/59)  & 53\%     (31/59)  \\ \hline
\textbf{3} & \multicolumn{2}{r|}{81\%     (25/31)}  & 55\%   (17/31) & 48\%     (15/31)  & 48\%     (15/31)  \\ \hline
\textbf{4} & \multicolumn{2}{r|}{64\%       (9/14)} & 71\%   (10/14) & 57\%       (8/14) & 43\%       (6/14) \\ \hline
\end{tabular}
\label{table:severity}
\end{table}

% \end{table}
%Next, we explain our findings about the performance of the synthetic evaluation. Recall that all sev = 0 issues are excluded from our finalized sets, including the synthetic evaluation set, expert evaluator set, and the master set. 

To evaluate the performance of the synthetic evaluation, we first contrasted its results against the individual experts' evaluation results. Excluded in this analysis is the set of issues in our master set that we coded as a severity of 0 (not a usability problem). We found that synthetic evaluation was able to identify significantly more usability violations than each of the individual expert evaluators. For the rental app, expert evaluators, on average, identified 18\% (24/133) problems. The synthetic evaluation, in contrast, uncovered 73\% (97/133) of the usability issues. For the language learning app, the expert evaluator on average identified 17\% (19.4/113) problems. The synthetic evaluation identified 77\% (87/113) of the usability issues. 

We also explored how synthetic evaluation compared against the aggregation of 5-expert evaluations. As prior research suggests~\cite{nielsen1990heuristic}, 3-5 expert evaluators provide sufficiently good coverage for all the usability problems (i.e., 55\%-90\%). We found that the synthetic evaluation achieved a higher coverage than the aggregation from our 5 expert evaluators for both apps. The aggregated 5-expert evaluation found 57\% (76/133) and 63\% (71/113) for the rental and language learning apps respectively, compared to the aforementioned 73\% and 77\% uncovered by the synthetic evaluation. See \autoref{fig:app1} and \autoref{fig:app2}. 

In addition, we found that synthetic evaluation reported more usability issues than the aggregated 5-expert evaluation for both the rental app ($p < 0.001$) and the language learning app ($p < 0.001$) across all severity ratings. This includes issues of severity 0, indicating that synthetic evaluation also found more non usability problems (\autoref{table:severity}).

We also compared the evaluation results across individual heuristics (\autoref{table:heuristic}). Again, synthetic evaluation consistently outperforms the aggregated 5-expert evaluation for most heuristics, especially in \textit{Aesthetic and minimalist design.} Synthetic evaluation achieved an 86\% (18/21) and 74\% (14/19) coverage for \textit{Aesthetic and minimalist design} for both apps, which is much better than the aggregated expert evaluators' coverage of 38\% (8/21) and 42\% (8/19). And synthetic evaluation is least effective in detecting violations of the heuristic \textit{Consistency and standards,} only covering 46\% (6/13) and 40\% (4/10) for the two apps tested, which is slightly worse than the aggregated expert evaluators' coverage of 54\% (7/13) and 60\% (6/10). Our analyses of issues identified, as we will report later, provide some potential explanations for these results.

Below we describe additional insights we gained from further analyses of our data. 
% \subsection{Synthetic evaluation performance is more consistent than human evaluation}
% Attention cost 
% number of issues
% limited by the number of issues one could identify
% Conversational grounding (time)
\subsubsection{Synthetic evaluation's performance remained consistent across tasks while human evaluation's performance decreased over time during the evaluation}
\begin{figure}[!tbp]
  \centering
      \begin{minipage}[b]{0.6\textwidth}
      \includegraphics[width=0.9\textwidth]{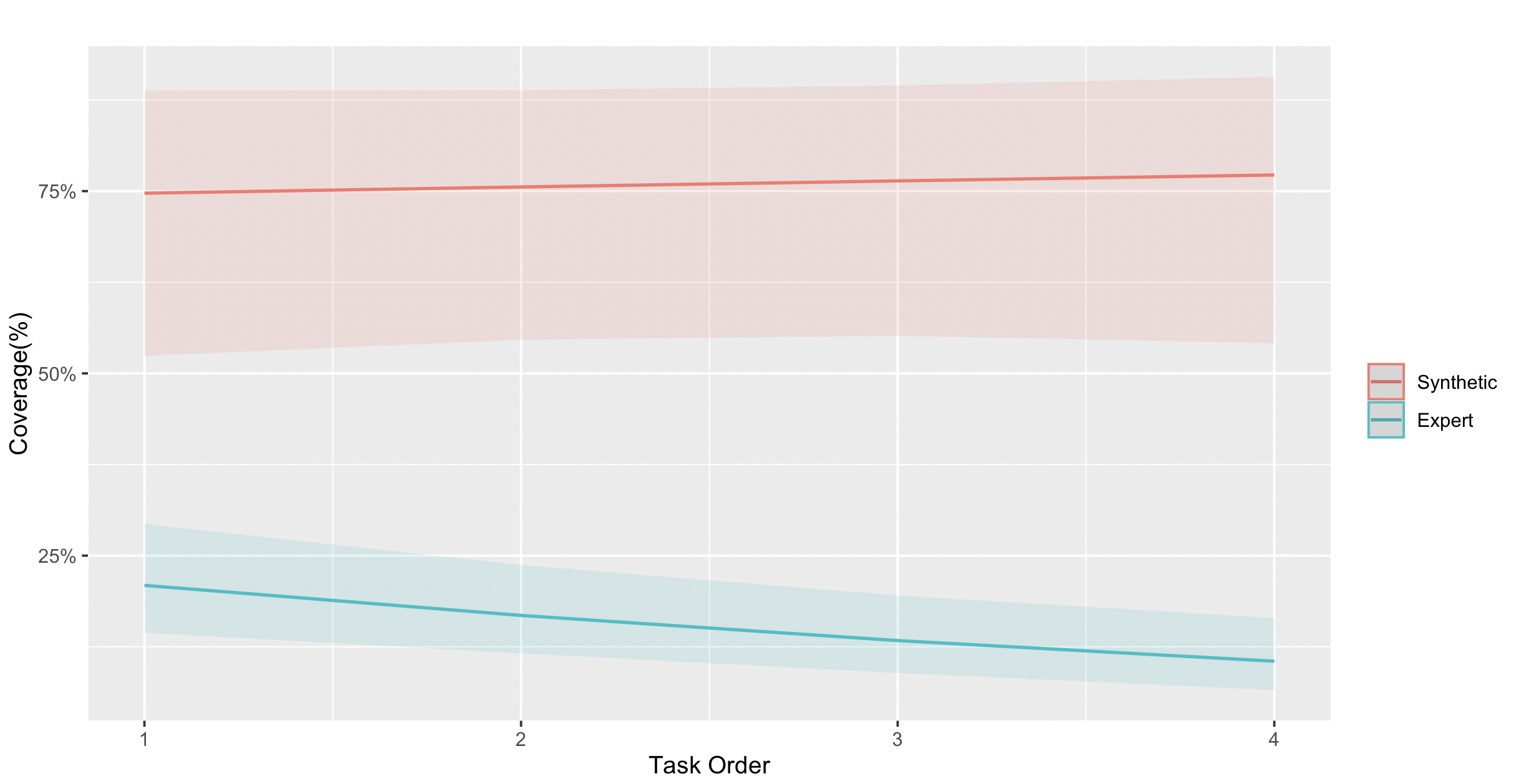}
    
  \end{minipage}
  \caption{This figure shows expert evaluators' performance and synthetic heuristic evaluation's performance across user tasks for the combined data of both the rental app and language learning app.}
    \label{fig:scenario}
  \hfill
\end{figure}

% TODO: mixed effect binomial logistic regression
% significant effect (p)
% for every increase in, there is an effect xxx
%  Our estimates suggest that the likelihood for a usability problem to be identified by synthetic review remained around 75\% across tasks. However, for human evaluators, the average likelihood starts off around 18\% but drops to about 10\% for the last task . 
Analyzing the performance of both synthetic evaluation and the expert human evaluation across evaluated user tasks, we found that while synthetic evaluation performance remained stable across tasks, human evaluation's performance decreased as they progressed from one evaluation tasks to the next (i.e., over the evaluation period). A mixed-effect binomial logistic regression shows that this interaction effect is significant ($p < 0.05$). As shown in \autoref{fig:scenario}, synthetic evaluation's results remain consistent while human evaluators' performance decreases by 31.6\% across the evaluation tasks ($OR = -0.316, 95\%CI [-0.60, -0.03]$). 

When analyzing their qualitative descriptions, we also observed a similar decrease in expressiveness, especially when similar problems showed up in later evaluation tasks than earlier ones. For instance, we found that the first time an aesthetic issue was reported, E3 would argue that \textit{``I do not see a lot of consistency in the design, especially in the colors.''} However, later on, when this issue was reported again, the evaluators would simply put, \textit{``more inconsistency in the use of colors.''} In contrast, synthetic evaluation consistently used the verbiage of \textit{``The use of coloring is inconsistent with the typical presentation of apartment features, which could confuse users.''} Similarly, in the second task of the language learning app, E6 pointed out that \textit{``screen 1,2, and 3 show a subtle change in the header which makes it difficult to understand what the screen changed to,''} whereas they only said \textit{``Pagination unclear''} the second time this occurred. Synthetic evaluation, however, used similar verbiage of \textit{``There is an unclear indication of the progress within the setup process. Users might not know how far they are from completing the initial setup accordingly.''} 

%As we will expand on further in the discussion, this may be because AI does not have the same cognitive limitations as human evaluators. Humans experience fatigue and have limited attention span, and thus may exert less effort and be less capable of identifying usability problems over time in a single sitting. On the other hand, synthetic evaluation, using compute power, is able to able to maintain a stable performance over time.

\subsubsection{Synthetic evaluation performed better in detecting small layout differences}
As discussed, synthetic evaluation performed better than the aggregated 5 expert evaluators regarding \textit{Aesthetic and minimalistic design} for both the rental app and the language learning app. Upon closer analysis, we found that the human evaluators did not identify as many issues about small layout differences as the synthetic evaluation. For instance, synthetic evaluation recognized that one of the screens in the language learning app is condensed with detailed information and that \textit{``the use of space could be optimized to provide more context or control features without cluttering the design,''} and thus reducing the cognitive load to users. Similarly, in the rental app, the synthetic evaluation picked up on small details, as shown in \autoref{fig:example1}, such as \textit{``the whitespace above and below the `Tour requested' text within the highlighted box is not consistent and too large, making it feel overly padded and overloading people''} and \textit{``there is only a little differentiation in font size between the `Tour' and `Tour requested,' leading to an unclear visual hierarchy.''}

\subsubsection{Synthetic evaluation had trouble recognizing and interpreting UIs}
\label{results:sev0}
While synesthetic evaluation identified more issues than human evaluators, our additional analyses showed that one aspect that it struggled with was recognizing and interpreting UIs. 

\paragraph{Difficulty of recognizing some UI components}
One limitation of the synthetic evaluation is that it sometimes misinterprets image components, leading to the identification of numerous issues that we determined to not be usability issues. For the rental app, 42\% (18/43) of the severity 0 issues identified by the synthetic evaluation were related to synthetic evaluation's inability to recognize specific interface components. For the language learning app, 27\% (14/52) of the issues identified by the synthetic evaluation were related to this aspect. For instance, the screenshots we used included information about the phone itself, such as the Wi-Fi signals or the battery level, as shown in \autoref{fig:example1}. The synthetic evaluation sometimes falsely considered these elements of the iOS system as part of the user interface. It would describe that these elements have usability issues: \textit{``The Wi-Fi icon in the status bar is not standard and could confuse users.''} Another example is where the synthetic evaluation was not able to correctly recognize a pop-up banner. Instead, the synthetic evaluation considered the element as a button and claimed that this banner \textit{``could be tapped accidentally without confirmation, possibly leading to unintended purchases or changes.''} Similarly, in the language learning app, there is a page asking participants to select their preferred time commitment where each option had a clock icon associated, indicating whether people should commit 5 min, 10 min, 15 min a day (\autoref{fig:example4}). Essentially, the clock icons and the options were presented together. However, synthetic evaluation failed to recognize that and argued \textit{``the clock icons could be interpreted as timers or alarms, which may not be relevant to the context of daily commitment.''} This example illustrates that synthetic evaluation still has difficulty understanding interface components. 
% However, because the icons are labeled with time intervials such as ``5 min / day'' or ``15 min / day,'' the case that the synthetic evaluation brought up is in fact highly unlikely. 
% And on a location searching screen where ``New York'' is shown as a suggested popular location in a selectable list, synthetic evaluation failed to understand that this is a  and said that \textit{``there is no indication that the `New York' section is clickable or that it expands into more options.'' }

% This example reflects that synthetic evaluation still has trouble understanding whether a specific component conforms to the guideline that icons should be clearly communicating the purposes. This finding illustrates that while expert evaluators could recognize how to determine whether a design principle is applicable to a 
% Among the 43 severity 0 issues identified by synthetic evaluation in the rental app, 21  Similarly, among the 53 severity 0 issues in the language learning app, 11 were related scenario, synthetic evaluation still has some trouble making that distinction. 

\paragraph{Lack of understanding of app conventions}
In addition, synthetic evaluation sometimes failed to understand the common conventions and/or design for mobile apps, again leading it to report certain non-issues. For the rental app, 42\% (18/43) of the severity 0 issues identified by the synthetic evaluation were related to some misunderstanding of design conventions. For the language learning app, 38\% (20/52) severity 0 issues identified by the synthetic evaluation were related to this. For example, in the rental app, there is a button that allows users to switch between a list view and a map view of rental listings as needed. This button, being a crucial function, was intentionally designed to stand out, using a distinct alerting color compared to other buttons so that people can easily use it to navigate. This is a common practice in rental apps. However, the synthetic evaluation did not recognize this convention, instead criticizing that \textit{``The `List' button to switch back to the list view attracts too much attention''} Similarly, on the setup page of the language learning app (\autoref{fig:example2}), where both a ``Choose a Language'' and a ``Log in'' button are displayed, the synthetic evaluation said that \textit{``The two buttons did not have the same design and the `Choose a language' button is much larger than the `Log in' button, which should not be prioritized. The `Log in' button should be designed as the main element.''} However, the design of the two buttons actually followed the button design convention in mobile apps~\cite{uxmobiledesign}. The ``Choose a language'' button uses a solid button design to indicate that this option would change the state of the app, changing the language setting. In contrast, the ``Log in'' button uses a linked button design to show that the screen will go to a different page where one could sign in instead of directly changing the state. The app's design choice aligns with common practices in mobile apps when handling login versus sign-up options. However, synthetic evaluation was unable to recognize that. Overall, these examples highlight synthetic evaluation's lack of understanding of standard app designs. This limitation may stem from insufficient training data related to user interfaces and mobile app conventions, suggesting that further fine-tuning is necessary to improve synthetic evaluation's performance in these areas.
% However, since this is the first page users see upon downloading the app, it is likely that most users do not yet have an account, making it reasonable to prioritize the ``Choose a Language'' button over the login option.
\subsubsection{Synthetic evaluation lacked the ability to aggregate information from multiple screens}
In addition, synthetic evaluation is still challenged to draw information from multiple screens to complete the evaluation. 

\paragraph{Synthetic evaluation repeatedly reported the same issues that occurred on multiple screens}
As we have reported, in our curation of the master set, we found that while the synthetic evaluation set sometimes contained duplicate issues (8 for the rental app and 9 for the language learning app), the expert human evaluators did not report duplicate issues. Exploring this, we noticed that these issues were all due to synthetic evaluation reporting on the same issues across different screens. For instance, in a series of screenshots showing the initial setup process in the rental app, the progress bar is small and confusing throughout the 5 screenshots showing the process. E3 said that \textit{``the progress bar does not seem to give complete information about how many steps there are remaining in the flow''} only on the first screen where the issue occurred. In contrast, the synthetic evaluation repeatedly said that \textit{``users might not know how far they are from completing the initial setup''} for each and every one of the 5 screenshots. Similarly, in the rental app, when users performed a search query, the hierarchy of the screen remained the same, with the filters at the top of the screen. However, as shown in \autoref{fig:example3}, there were too many filters presented, which was difficult for people to navigate. While E2 only pointed out that \textit{``there are too many filters, users need to scroll horizontally to view all filters''} on the first screen, the synthetic evaluation repeatedly discussed this issue for all screens that had the filters at the top. This may be because synthetic evaluations consider each screen as its own separate artifact and do not consider whether an issue has already been identified regarding the same component on a previous screen. We may be able to improve this aspect by trying out more prompting techniques to encourage the synthetic evaluation to only identify new issues instead of repeatedly identifying the same issue.  
% For example, we may be able to integrate chain-of-thought prompting to enhance LLMs' memorization of previous screen information~\cite{achiam2023gpt}. 
% First of all, we observed that synthetic evaluation would consistently report the same issues that occurred on multiple screens repeatedly. 
% In the rental app, the synthetic evaluation identified 8 such issues. In the language app, synthetic evaluation pointed out 9 such issues. In contrast, the expert evaluators did not do this type of repeated reporting. 

\paragraph{Synthetic evaluation had some trouble identifying across screen violations}
We also found that the synthetic evaluation had difficulty identifying violations that involved multiple screens. We noticed that synthetic evaluation only found 43\% (3/7) of the across screen violations for the rental app, whereas the expert evaluators uncovered 86\% (6/7). In the language learning app, the synthetic evaluation discovered 50\% (3/6) of the across screen violations, whereas the expert evaluators uncovered 83\% (5/6). For instance, in the rental app, E4 noticed that the layout of one of the screens is completely different from the other ones in a continuous setup process, and thus called out that \textit{``a full screen and blurred background is not expected, which is inconsistent with the other screens.''} Similarly, in the language learning app, E7 pointed out that \textit{``the use of terms such as courses, collections, lessons, and sessions is inconsistent''} on multiple screens, where each screen was using one of these terms. The synthetic evaluation may have missed this violation because it did not understand that the screenshots given all belong to the same setup process. This finding also corresponded with our earlier observation that synthetic evaluation performed worse than the expert evaluators regarding the heuristic \textit{Consistency and standard}, because all the across screen violations that synthetic evaluation missed in these two apps are violations of this heuristic.
\section{Study 2: Synthetic Evaluation Reliability \& Cross-Platform Testing}
Next, we conducted a study to test the performance of synthetic evaluation with repeated prompting across time and accounts. In addition, we compared the performance of synthetic evaluation across multiple platforms.

\subsection{Procedure}
\subsubsection{Across Time \& Accounts}
We tested the reliability of our synthetic heuristic evaluation to see if using it across time (over a three-month period, March 20th - Jun 17th 2024) and across two accounts would result in a consistent set of issues. While the synthetic evaluation output of heuristic violations was not exactly the same each time, the main difference was in the wording. Different descriptions would still point to the same heuristic violation. For instance, \textit{``no progress indication,''} \textit{``lack of onboarding or progress indicator,'' }and \textit{``no indicator pointing out where users are in the setup process''} all refer to the same issue. The first author coded the data from multiple time periods and accounts based on these criteria. Two other researchers cross-checked the labeling and discussed any discrepancies. 

\subsubsection{Across Platform}
We also contrasted the synthetic evaluation output across multiple platforms: GPT-4, Gemini-1.5-pro, and Claude 3.5 Sonnet. We use the same synthetic set from \autoref{synthetic-set} for the GPT-4 analysis. For Gemini-1.5-pro and Claude 3.5 Sonnet, we used the same prompt that we did with GPT-4 and collected their outputs. 

To analyze the data, we followed a similar procedure as the procedure discussed in \autoref{section:masterset}. We combined all the heuristic evaluations found by Gemini-1.5-pro and Claude 3.5 Sonnet with the master set. If the two models detected additional issues that were not detected in the original master set, we would add it in. 

The first author coded all the data from Gemini and Claude, and two other researchers cross-checked the work to confirm the codes. We found that the two models did not uncover any new issues that are not of severity 0 for the two apps. Therefore, the master set remained the same. But we did discover that these two models found some additional issues that are of severity 0. For the rental app, Gemini discovered 9 and Claude found 12 additional severity 0 problems. For the language learning app, Gemini found 13 and Claude found 9 additional severity = 0 issues. Therefore, we added these issues to our compiled set: the total of the severity 0 set for the rental app is 70 (49 from the analysis in \autoref{section:masterset}, 21 from Gemini and Claude); the total of the severity 0 set for the rental app is 77 (55 from the analysis in \autoref{section:masterset}, 22 from Gemini and Claude) Details of the performance of Gemini and Claude are reflected in \autoref{table:heuristic} and \autoref{table:severity}.

\subsection{Analyses}
For analyses of the synthetic evaluation data across time and accounts, we calculated both coverage-consistency (i.e., against the first set of results) and performance-consistency (i.e., against the master set). For analyses of the across platform data, we calculated each model's performance against the master set and compared the set of issues they detected.

% The first author labeled each issue based on whether the descriptions point to the same issue. Two other researchers cross-checked the result and discussed any discrepancies. If any of the models detected additional issues that were not detected in the original master set, we would add it in.

\subsection{Results}
\label{reliability-result}
\subsubsection{Across Time \& Accounts}
First, we tested GPT-4's coverage-consistency (i.e., against the first set of results). We found that for the rental app, both accounts tested were able to detect 94\% (171/182) of that set. Similarly, for the language learning app, account 1 identified 95\% (156/164), and account 2 identified 96\% (157/164). And over three months, GPT-4 achieved high coverage-consistency for the rental app ($M = 93.86\%, SD = 0.016$) and the language learning app ($M = 93.90\%, SD = 0.021$). Next, we tested to make sure GPT-4 performed consistently well in identifying the heuristic issues. We found that for the rental app, account 1 identified 70\% (93/133) and account 2 identified 71\% (94/133) of the master set. For the language learning app, both accounts identified 77\% (87/113) of the master set. And over three months, GPT-4 consistently achieved a similar performance for both the rental app ($M = 69.55\%, SD = 0.016$) and the language learning app ($M = 74.85\%, SD = 0.015$). These results demonstrate that our synthetic evaluation approach is consistent with repeated prompting.

\subsubsection{Across Platform}
We also contrasted GPT-4's performance against that of Gemini-1.5-pro and Claude 3.5 Sonnet. We found that for the rental app, Gemini identified 59\% (79/133), and Claude identified 57\% (76/133), which is lower than GPT-4's performance of 73\% (97/133). For the language learning app, Gemini found 59\% (67/113) and Claude identified 58\% (65/113), which is lower than GPT-4's performance of 78\% (88/113). This shows that GPT-4 achieved the best coverage among all models tested, and Claude had the lowest. In addition, the performance of the Gemini and Claude model was about the same as the aggregated 5 human evaluators for the rental app at 57\% (76/133) and slightly lower for the language learning app at 63\% (71/113), as shown in \autoref{fig:app1} and \autoref{fig:app2}. 

Then, we compared the evaluation results across individual heuristics (\autoref{table:heuristic}). We noticed that GPT-4 found more usability issues than Gemini and Claude for severity ratings 1 through 4. For severity 0, Gemini and Claude reported more issues than GPT-4, though not by a significant amount. 

We also contrasted the evaluation results across individual heuristics (\autoref{table:severity}). Again, GPT-4 performed better than Gemini and Claude for all heuristics, especially in \textit{Visiblity of system status} and \textit{Match between system and the real world}. After comparing the set of issues identified by all three models, we found that the issues that Gemini and Claude failed to find in comparison with GPT-4, are issues related to understanding the app conventions and identifying issues that violate these conventions. As specified in \autoref{results:sev0}, GPT-4 also struggled with this issue. However, GPT-4 was able to find more of this type of issue than Gemini and Claude, leading to a better performance than the other two models. For instance, for the rental app, on a screen where multiple filters are present for a rental search query, GPT-4 pointed out that \textit{``The labels and icons used for the filters are not explicit about what they filter, which could lead to a mismatch between the user's expectations and the system's operations.''} Identifying this issue requires identification of the UI elements and an understanding of the context of the rental application as well as what terminologies are most appropriate. But Gemini and Claude were not able to do so and identify this violation. Similarly, for the language learning app, during the lesson setup process, the screen shows irrelevant statistical information about the app. GPT-4 directly pointed out that\textit{ ``The statement about the percentage of beginners doesn't provide actionable information for the user's personal journey,''} demonstrating an understanding of what a setup process is and what information should not be involved in this process. Again, Gemini and Claude did not find this violation. Therefore, while it is true based on our previous discussion that GPT-4 struggles to understand some app conventions, it is still doing better than Gemini and Claude in this aspect.

% For example, for the rental app, on a screen where there was no heading provided to demonstrate the status of the search process, GPT-4 identified that \textit{``there was no heading available for the screen, which can be confusing for users,''} whereas Gemini and Claude did not find this issue.   

% An analysis of the severity 0 issues that Gemini and Claude found reveals that they also have difficulty in recognizing some UI components and a lack of understanding of app conventions, which are the same challenges that GPT-4 face, as mentioned in \autoref{results:sev0}. 
% For instance,  
% In addition, for the rental app, when the user needs to select a timeframe to show listings available during that timeframe, the app only allowed for single choice selection because the timeframes . But Gemini mentioned that \textit{``users can't select multiple time periods''} and Claude said that \textit{``users are not allowed to choose multiple options.''} These comments indicate that Gemini and Claude, similar to GPT-4, does not understand app conventions and the context of a piece of UI design.

\section{Discussion}
Our work explored the use of multimodal large language models (LLMs) to conduct heuristic evaluation and compared its performance against that of expert human evaluators. We found that the synthetic evaluation achieved higher coverage of the master set of usability issues compared to the aggregated performance of five expert evaluators across two apps. 
% This is the first systematic, empirical demonstration that current off-the-shelf LLMs can be effectively used for heuristic evaluation. 
In addition, we found that our synthetic evaluation approach was able to produce assessments of interfaces leveraging the given set of heuristics. Unlike existing automated usability approaches that produce primarily quantitative scores that are hard to interpret~\cite{baumeister2000comparison,glenn1992development,al1999kaldi,hammontree1992integrated,fosco2020predicting,lee2020guicomp,uehling1995user}, our approach also presented qualitative descriptions discussing potential user preferences, similar to human evaluator outputs. As pointed out by Ivory \& Hearst~\cite{ivory2001state}, this type of subjective data is a key aspect of usability testing and has been a missing piece in existing automated usability testing systems. Moreover, while prior work only demonstrated LLM's capability to produce design suggestions~\cite{wu2024uiclip,duan2024uicrit,duan2024generating}, we showed its application in conducting a usability testing method -- heuristic evaluation, which requires more reasoning ability to explain how each issue violates a heuristic. Our results indicate that LLMs are able to analyze images, draw on information about similar designs, compare the design against a given set of heuristics, and identify issues that violate the heuristics.

% Additionally, using multimodal LLM, our approach is able to better support automated usability testing. Specifically, our approach only needed to provide screenshots of the user interfaces and short descriptions of the user tasks to the LLM -- materials that are currently shared with evaluators. In contrast, prior systems required the use of text-based descriptions (e.g., JSON descriptions) of the screens as inputs~\cite{duan2024generating}, or conducting testing sessions with actual users~\cite{rauterberg1995novice,cugini1999visvip,webtrends}. Gathering such additional information can be costly and difficult for those conducting evaluations, especially if they do not have the expertise to produce these descriptions or collect the testing sessions. Thus, our approach may be more easily integrated into existing evaluation workflows and lower the barrier for people to conduct testing~\cite{usabilitycost,userevalstats}. 

Further, our work demonstrated the reliability and generalizability of using generative AI to conduct synthetic evaluation. Recently, scholars and practitioners~\cite{morris2023scientists,staudinger2024reproducibility} have started to pay attention to testing the reliability of using LLMs, such as understanding how the stochastic nature of LLMs may impact the replicability of the results. If the model can only achieve high performance occasionally, it would be challenging to implement the approach at a large scale. While reliability has been tested in many applications~\cite{chang2024survey}, existing systems~\cite{liu2023chatting,duan2024generating, wu2024uiclip,duan2024uicrit} that use generative AI to generate design feedback have not considered this aspect. Our work demonstrated the consistency of the LLM across repeated prompting (over time and across accounts), highlighting the scalability of synthetic heuristic evaluation. Future work using LLM and/or testing LLM's ability in design should also conduct tests around reliability to ensure that the results can be replicated. 

Our work also tested the performance of multiple platforms/models (GPT-4, Gemini-1.5-pro, Claude 3.5 Sonnet) in conducting heuristic evaluation. We found that, generally, the AI-powered evaluations achieved better or similar performance than the aggregated expert human evaluators, which indicates that the synthetic evaluation approach may be generalizable across off-the-shelf LLMs. Among the three models tested, GPT-4 achieved the highest performance, and Claude achieved the lowest. In addition, we found that though all models were challenged to understand app conventions, GPT-4 did the best among the three. This may be because we did not change the prompt when testing the performance, which indicates the potential need to fine-tune prompting when using a different model. This may also be related to how each model is trained and the type of data used during training. Future work can conduct experiments to more closely investigate these model differences and explore how to fine-tune each model for better performance. 
% since we did not change the prompting as we tested across platform, further fine-tuning may be needed 

% In addition to demonstrating the feasibility and efficacy of using LLMs to conduct heuristic evaluation, our work also pointed out several benefits and drawbacks when using synthetic heuristic evaluation outputs in practice.

\subsection{Benefits and Challenges of Designing with LLMs for Heuristic Evaluation}
Recently, there has been growing interest in using AI to simulate users to support evaluation and research~\cite{duan2024generating, wu2024uiclip, liu2023chatting}. For instance, Synthetic Users is an AI-powered tool that creates a synthetic user persona and provides feedback according to that profile. Duan et al.~\cite{duan2024generating} and Wu et al.~\cite{wu2024uiclip} also developed AI-driven approaches to provide design feedback. Yet prior works did not explicitly compare LLM outputs against human evaluation and analyze the differences between the two. We address this gap in the existing literature and show insights about the differences in performance. These insights demonstrate what people need to know about synthetic evaluation when using them and how LLMs and humans would approach the same task differently. Moreover, these benefits and challenges point out key aspects to consider when designing LLM-powered usability evaluation tools.

% \subsection{Contrasting Synthetic Evaluation and Human Evaluation}
% Recently, there has been growing interest in using AI to simulate users to support evaluation and research~\cite{duan2024generating, wu2024uiclip, liu2023chatting}. Though our work is specific to heuristic evaluation, we found some interesting insights about the differences between human and synthetic evaluation that may be broadly relevant in all the work seeking to use LLMs to emulate humans in research and design. 

In terms of benefits, we found that the performance of synthetic evaluation remained consistent, while human evaluators' performances decreased over time. One likely explanation is that there are attentional differences between the LLM and human evaluators. Specifically, people's ability to process new information and make decisions is limited~\cite{chandler1991congnitive,sweller1988cognitive}. During a cognitive intensive task such as evaluation, where evaluators need to pay attention to the design as well as make judgments about them, humans often experience cognitive overload towards the end~\cite{tracy2006measuring}. Prior works have discussed that this cognitive load may lead to human evaluators only doing enough work and stop when they feel satisfied~\cite {american1966theories, redish2007expanding}. This would explain the observed decrease in human evaluators' performance as they progress through the tasks. In contrast, the synthetic evaluation does not experience this attentional constraint because LLMs’ performance only relies on compute power. The difference between how LLM and human function highlights the benefit of using generative AI for synthetic evaluation. Noting this characteristic, when using synthetic evaluation, practitioners do not need to worry about the number of tasks tested and the complexity of the tasks, which is a great solution to the rising cost of running heuristic evaluations with humans~\cite{usabilitycost}. 

Our results also demonstrate that synthetic evaluation was better at detecting small layout differences than human evaluators, potentially highlighting another key benefit of LLMs. It is well known that humans have difficulty detecting small differences in visuals due to physiological and psychophysical mechanisms in humans' visual system ~\cite{zhang2008just, lin2005visual}. However, since LLMs are built to read images through encodings and rely on feature extraction~\cite{achiam2023gpt}, it is likely that the models are particularly good at recognizing small differences in layout such as font size consistencies and how cluttered an interface is~\cite{cao2024visual,fleuret2011comparing}. While more work is needed to further uncover the underlying mechanism of how synthetic evaluation interprets images, our findings suggest another potential difference and benefit of using synthetic evaluation. When reviewing synthetic evaluation outputs, practitioners should also consider layout issues potentially as of higher quality than other types of issues. This will help them reduce the amount of brainpower needed to process every single suggestion made. 

In terms of challenges, an important one to consider is that practitioners need to handle output processing when using synthetic evaluation. The generated outputs from LLM, as we discussed, contained duplicates of the same issue that occurred repeatedly across multiple screens. Thus, posthoc labeling and integration of the repeated issues is required. Additionally, due to its stochastic nature, every time the LLM is prompted, it may generate different outputs. But as discussed in \autoref{reliability-result}, the different descriptions may point to the same heuristic violation since the LLM is relatively stable when conducting synthetic evaluation. Therefore, when running an AI-powered heuristic evaluation multiple times, practitioners need to be aware and compare the generated descriptions and identify whether they point to the same issue. More importantly, our results indicate that LLMs identified many non-issues (severity = 0), which are not actual heuristic evaluations from practitioners' standpoints. To utilize LLMs' outputs, practitioners need to develop an efficient way of processing the large set of heuristic violations identified by LLMs and remove all non-issues. This may be possible with more training data of existing heuristic evaluations. However, some manual labor may still be required to ensure the quality of synthetic evaluation. 

In addition, we found that synthetic evaluation had some trouble correctly recognizing interface components and understanding app conventions. For instance, our approach was not able to correctly differentiate iOS components from the interface sometimes (\autoref{fig:example1}). And it did not understand some of the common design conventions in rental apps and language learning apps (\autoref{fig:example2}). This is likely due to a lack of interface-focused training data with current multimodal LLMs. We believe this difference may narrow as multimodal LLMs are trained with more images of technology interfaces, showing common designs used in mobile applications~\cite{ahmed2022few, song2023llm}. Thus, when practitioners utilize off-the-shelf LLMs, they need to be aware of such limitations, and fine-tune the model with more data. If resources are limited in doing so, they should at least be aware of this issue and pay close attention when LLMs point out violations related to interface understanding and app convention knowledge.

Our findings also suggest that synthetic evaluation lacks the ability to integrate information from multiple screens. We believe that this may be at least partially due to our prompting, which may be improved. Even though we have noted this as a problem already in our initial iterations in \autoref{section:prompt}, our approach might still be lacking. Our findings suggest that future versions may consider explicit prompts to have the synthetic review to filter out some of this duplication. Part of the issue is also that LLMs have limited memories, and future work could consider approaches such as RAG (Retrieval-Augmented Generation)~\cite{gao2023retrieval} to ensure that all information throughout the evaluation (including screenshots of user interactions, instructions, etc.) is accessible to LLM, improving the performance. Again, when partitioners utilize LLM-powered heuristic evaluation, they should pay close attention to these types of issues to make sure that all potential usability issues are identified.

Overall, our work highlights both the feasibility and promising potential of synthetic evaluation. Our findings show that it performed better than the aggregated 5 expert evaluators and offers additional benefits of detecting small differences in visuals and not suffering from attentional cost throughout. Using LLMs, we are now able to conduct a high-quality, large-scale heuristic evaluation at a very low cost. Yet there are specific characteristics of LLM outputs that are different from humans, offering an opportunity for us to rethink AI-powered heuristic evaluation tool designs. We should think about how to best organize and work with LLM outputs to ensure that: 1) the effective and high-quality information in the synthetic evaluation is extracted and used efficiently 2) the potential non-issues are all identified so that practitioners do not spend efforts identifying and fixing them 3) the areas where synthetic evaluation's performance is unstable are closely checked to ensure quality of the result. While existing tools have explored using AI on the backend to provide design critiques~\cite{duan2024generating}, they have not considered how to most effectively convey AI outputs to the users. In our paper, we start a conversation about areas one should consider in design to answer that question. And our results point out that it is now time to reflect on more than just the feasibility of AI. We should also focus on what AI is generating and how we could most effectively leverage AI outputs to help the targeted users in design.

% As multimodal LLMs continue to improve and as we figure out more effective prompting techniques, the number of duplicates and non-issues reported by synthetic evaluation should also decrease. the question about the use of synthetic evaluation extends beyond job displacement, and future research should identify the long-term impacts of using synthetic evaluation and offer insights on how to effectively integrate the use of synthetic and human evaluation. 

% The question is then no longer whether we can but rather whether we ought to be using synthetic evaluation. From a job displacement perspective, at least, our proposed approach is unlikely to diminish the role of the researcher. While synthetic evaluation is likely to reduce evaluation jobs for \textit{e.g.}, those in online human subject pools such as Prolific and MTurk, trained researchers may be more in demand as they will be needed to help analyze the growing set of evaluation results if AI is going to enable more people (including non-experts) to conduct more evaluations more frequently. But 
\section{Limitation}
Our study only tested synthetic evaluation's performance on two types of commonly used apps. Further research may explore how our findings may apply to more domain-specific apps that require knowledge about targeted populations. Additionally, as generative AI models continue to evolve over time, their performance may change accordingly. Although we tested the reliability of GPT-4 across accounts and over a three-month period, future refinements to the model could impact its performance. Consequently, future applications of generative AI for synthetic evaluation may require adaptations in prompting techniques to maintain effectiveness.
\section{Conclusion}
This paper explores the potential of synthetic heuristic evaluations, presenting characteristics of synthetic evaluation compared to human evaluation results. Our findings indicate that synthetic heuristic evaluations outperform the average human evaluator and can achieve better performance than that of five human evaluators. Our work also underscores the importance of testing the reliability of LLM outputs when assessing their performance. Overall, this study demonstrates a promising approach to reducing the cost of human evaluations in human-centered design and enhancing automated usability testing processes.
\bibliographystyle{ACM-Reference-Format}
\bibliography{10-bibliography}

\appendix

\end{document}